\documentclass[letter]{aa} 

\usepackage{graphicx}
\begin{document}

   \title{Dynamical evolution of a self-gravitating planetesimal disk in the 
distant trans-Neptunian region}

   \author{V.V.Emel'yanenko
          \inst{}}

   \institute{Institute of Astronomy, Russian Academy of Sciences,
              Moscow, 119017 Russia\\
              \email{vvemel@inasan.ru} }

   \date{Received                       ; accepted                          }

 
  \abstract
  {}  
   {We study the dynamical evolution of a system consisting of the giant planets and a massive planetesimal disk over the age of the Solar System. The main question addressed in this study is whether distant trans-Neptunian objects could have come about as a result of the combined action of planetary perturbations and the self-gravity of the disk.}
   {We carried out a series of full N-body numerical simulations of gravitational interactions between the giant planets and a massive outer disk of planetesimals.}
   {Our simulations show that the collective gravity of the giant planets and massive planetesimals produces distant trans-Neptunian objects across a wide range of the initial disk mass. The majority of objects that survive up through the age of the Solar System have perihelion distances of  $q>40$ au. In this region, there is a tendency toward a slow decrease in eccentricities and an increase in perihelion distances for objects with semimajor axes $a>150$ au.  Secular resonances between distant planetesimals play a major role in increasing their perihelion distances. This explains the origin of Sedna-type objects. In our integrations for the age of the Solar System, we registered times with both high and low clustering of longitudes of perihelion and arguments of perihelion for objects  with $q>40$ au,  $a>150$ au. The resulting distribution of inclinations in our model and the observed distribution of inclinations for distant trans-Neptunian objects have similar average values of around $20^\circ$.}
   {Distant trans-Neptunian objects are a natural consequence in the models that include migrating giant planets and 
a self-gravitating planetesimal disk.}

\titlerunning{ A self-gravitating disk of distant trans-Neptunian objects}
  \maketitle

%

\section{Introduction}

\begin{figure}[h]
   \centering
      \resizebox{\hsize}{!}{\includegraphics{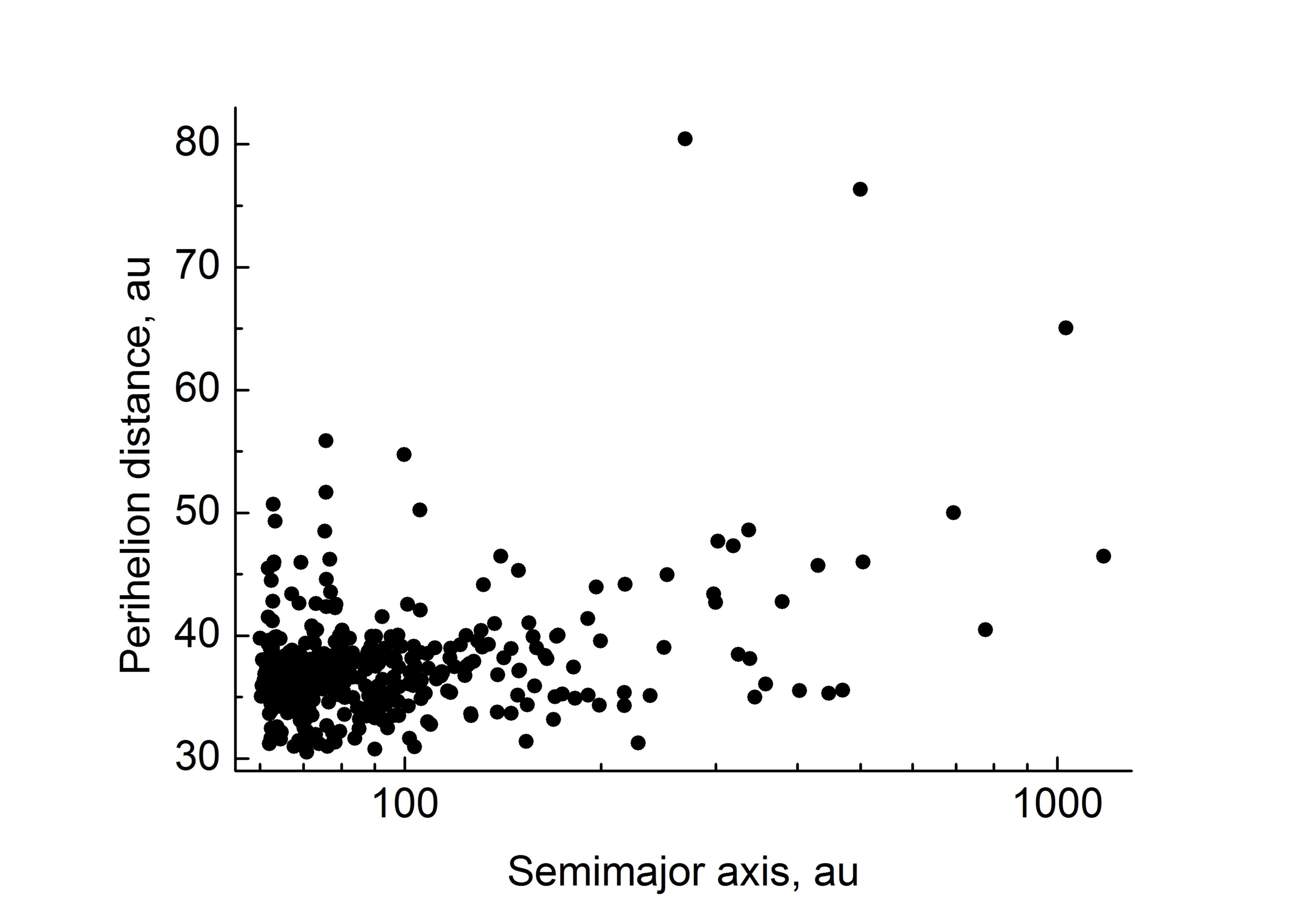}}
      \caption{Distribution of semimajor axes and perihelion distances for observed 
      multiple-opposition TNOs with $q>30$ au and $a>60$ au.}
         \label{fig1}
   \end{figure}

The population of trans-Neptunian objects (TNOs) moving in orbits with large eccentricities exhibits a complicated orbital structure. Figure~\ref{fig1} shows the distribution of semimajor axes, $a,$ and perihelion distances, $q,$  in the region of  $q>30$ au, $a>60$ au for TNOs observed in at least two oppositions (the data are taken from the Minor Planet Center database\footnote[1] {https://www.minorplanetcenter.net/iau/lists/Centaurs.html} on January 9, 2022). Objects with perihelia near the orbit of Neptune experience large perturbations from this planet. These objects are widely believed to have originated through scattering from initial orbits located close to Neptune and, hence, they have been referred to as "scattered disk objects" \citep{Luetal97,DL97}. On the other hand, most of the observed objects with $q>38 $ au, $a>50$ au  cannot be explained by a model based on a near-Neptune disk of planetesimals that is gravitationally scattered by Neptune \citep{Emetal03}. It has been shown in a number of works \citep[e.g.,][]{Go03,KS16,Neetal16} that orbits with large perihelion distances can arise during the migration of Neptune in a massive disk of planetesimals. This scenario is associated with a secular increase in perihelion distances for planetesimals moving in mean motion resonance with Neptune.  \citet{KS16} and \citet{Neetal16} noted that the distribution of orbits with $q > 40$ au in their studies has a dominant concentration near Neptune's mean motion resonances $s:1$.  Using the data for $s \le 13$, \citet{KS16} found that  the produced high-perihelion populations fall off as $s^{-1.4}$. \citet{Goetal05}, \citet{Saetal17}, and \citet{CS21} showed that orbits with $q > 40$ au can occur even for larger values of $s$. However, this mechanism fails to explain the origin of objects with extremely large perihelion distances, such as Sedna and 2012 VP113  \citep[e.g.,][]{Saetal17}.
 
As was shown by \citet{MM16}, in solving the problem of the origin of distant objects,  the self-gravity of the planetesimal disk can play a very important role. The authors of this paper found an effect of systematic changes in orbital characteristics driven by the collective gravity of bodies moving in orbits with large eccentricities, which they called  "inclination instability." This dynamical instability exponentially grows the orbital inclination, $i,$ of bodies while decreasing their orbital eccentricities, $e,$ and clustering their arguments of perihelion, $\omega$. Various features of this phenomenon were considered in a range of works \citep{Maetal18,Fletal20,Zdetal20, Zdetal21}, including the possible development of the apsidal clustering of orbits. \citet{ZM20} studied the dynamical evolution of two systems with additional consideration of planetary perturbations, one of which was intended to reflect the orbital features of the scattered disk, and showed the possibility of forming orbits with very large perihelion distances ($q >100$ au).

In the model studied by \citet{ZM20}, the initial orbits had a fixed value of the perihelion distance ($q=30$ au) and planetary migration was not taken into account. Therefore, it remains unclear whether the population of distant TNOs could have formed in the same process as other TNOs moving in orbits with large eccentricities. In particular, several works \citep{Go03,KS16,Neetal16} successfully explained the origin of a population of TNOs with $q>40$ au,  $50<a<100$ au, and large inclinations ($i>20^\circ$) during Neptune's migration in the massive disk of planetesimals. However, these papers did not take into account any self-gravity among the planetesimals.

In this letter, we consider a dynamical process similar to that considered in the works \citep{KS16,Neetal16}, but taking the gravitational interaction of planetesimals into account. In addition, while \citet{KS16} and \citet{Neetal16}  mainly investigated  the origin of objects with semimajor axes $ 50<a<100$ au, we are interested in studying the dynamical processes that lead to the transition of TNOs to orbits with $a>150$ au. The main question addressed in this study is whether distant TNOs could have arisen due to the combined action of planetary perturbations and self-gravity of the outer planetesimal disk with a mass in the range that is currently accepted in models of the outer Solar System formation \citep[e.g.,][]{Tsetal05,Moetal07,BB10,Leetal11,NM12,Reetal15}. We note that \citet{FB17} and \citet{ST19} doubt that the dynamical effect found in \citep{MM16} could lead to the observed features of the orbits of distant TNOs.
  
\section{Methods}

We study a system of bodies consisting of four giant planets with their present-day mass and 1000 planetesimals, of which 170 planetesimals are massive and have the same mass, and the rest have zero mass. We consider models in which the total mass of planetesimals $M_d$  has different values: 5, 10, 15, 20, 25, 30, 35, and 40 $M_E$, where $M_E$ is the mass of the Earth.

The initial conditions for the bodies are close to those used in the models \citep{KS16,Neetal16}. The planets begin to move in orbits with small eccentricities ($e<0.05$) and orbital inclinations ($i<2.5^\circ$). The initial values of semimajor axes are 5.3, 9.1, 17.2, and 24.0 au for Jupiter, Saturn, Uranus, and Neptune, respectively. Semimajor axes of planetesimals are distributed according to the law $a^{-0.5}$ between 24 au and 30 au, and eccentricities and orbital inclinations are uniformly distributed in the intervals (0, 0.01) and ($0^\circ$, $0.5^\circ$), respectively. Mean anomalies, longitudes of ascending node, and arguments of perihelion are drawn from a uniform distribution in (0, $360^\circ$).

We carried out the full N-body simulations. Orbits were integrated for 4 Gyr, using the symplectic integrator \citep{Em07}.  For each object $k$, the time-step of the integrator is approximately equal to $25 r (1+4 B)/\varphi$ days, where $\varphi = 1+ B r + \gamma r  \sum_{j=1 (j \ne k)}^{174} \sqrt{m_j 
(m+m_j)}/{\Delta_j^2} + \gamma_1/r^{3/2}$; $r$ and $m$ are the heliocentric distance and the mass of the object $k$; $m_j$ is the mass of the perturbing object $j$; $\Delta_j$ is the distance between the object $k$ and the perturbing object $j$; 
$\gamma=5550$ and $\gamma_1 =8$ (we used the astronomical system of units). The small positive constant B was chosen so that the integration step never exceeds 450 days.

We removed objects from the basic simulations if  $a>1000$ au   (5000 au in additional variants) or $e>1$ far away from perturbing bodies. We stopped the integrations if the planetary system became unstable (see discussion below). We also removed objects that come within 0.005 au of the Sun.

A direct N-body integraton is computationally very expensive. We find that the choice  of (170+4) massive objects is a good compromise when working to integrate over the age of the Solar System. In this case, each 4 Gyr run takes  months of CPU time.  

\section{Results}

\subsection{Basic integrations}
 
It is known from numerous works on planetary migration in planetesimal disks that four-planet systems are typically destabilized \citep[e.g.,][]{Moetal07,BB10,Leetal11,NM12,Reetal15,FB17,QK19}. In this study, we looked for specific variants in which all four giant planets remain in almost circular orbits with small inclinations for the age of the Solar System. Here, we discuss such examples that we found for $M_d =15M_E$, $M_d =20M_E$, $M_d =30M_E$, and $M_d =40M_E$. After running hundreds of  simulations with random initial orbits of planetesimals, we did not find any variant with $M_d <15M_E$ in which the planetary system is stable for 4 Gyr. 

The migration rates of the planets and their final positions are different in these variants. Therefore, in order to compare the results of our simulations with the observed orbital distribution of  distant TNOs, we introduced the parameters    $a^*=a(30/a_N)$ and $q^*=q(30/a_N)$, where $a_ N$ is the semimajor axis of Neptune at a given moment of time.  Neptune's semimajor axis evolution is shown in Fig.~\ref{figa} in Appendix A. The evolution of disk masses is shown in Fig.~\ref{figb} in Appendix B.

\begin{figure}[h]
   \centering
   \resizebox{\hsize}{!}{\includegraphics{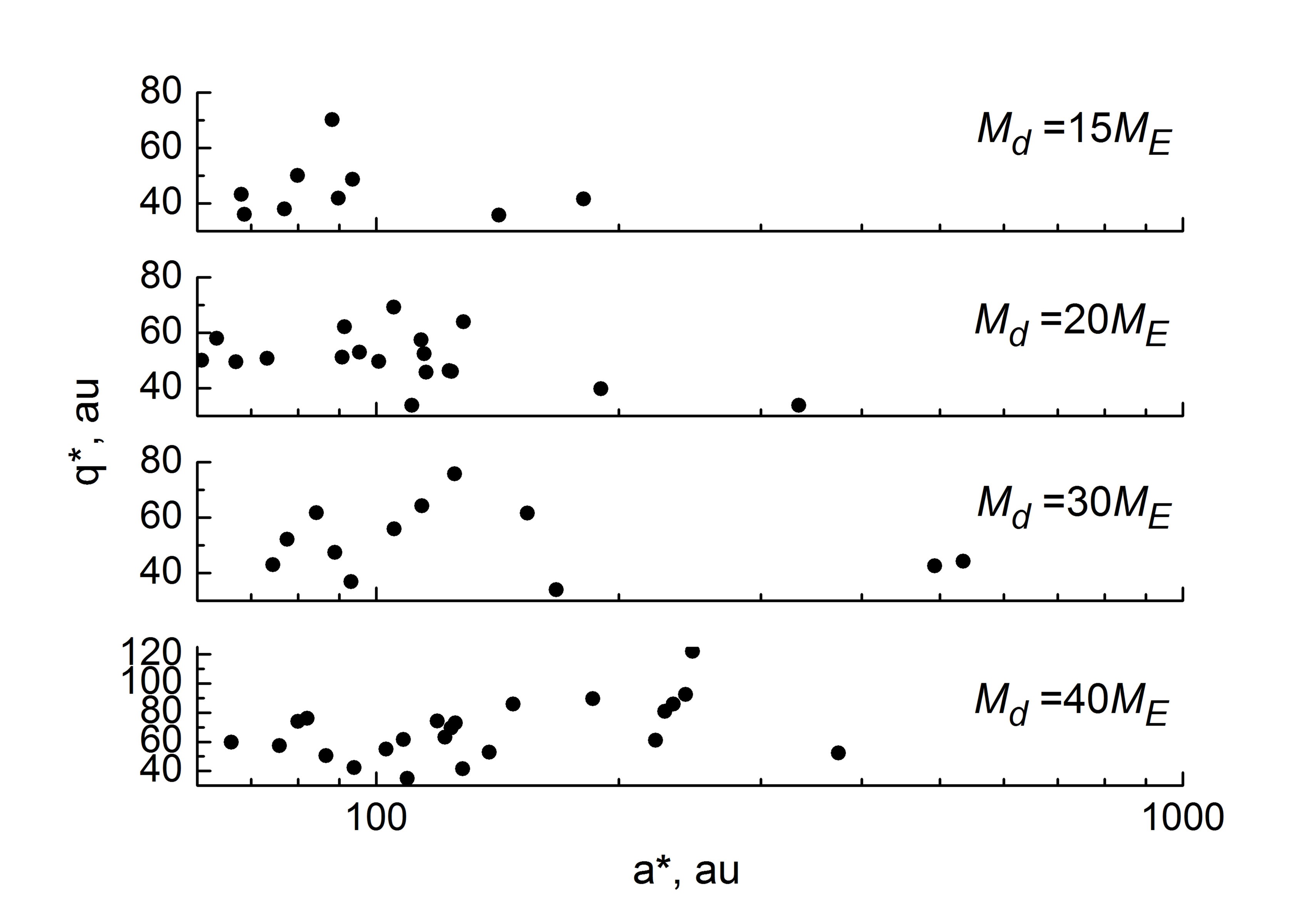}}
      \caption{Distribution of $a^*$ and $q^*$ after 4 Gyr for various values of  $M_d$.}
         \label{fig2}
   \end{figure}

 In all the stable cases, the majority of objects that survive for the age of the Solar System have   $q^*>40$ au. Figure~\ref{fig2} shows  the distribution of $a^*$ and $q^*$ for objects (planetesimals) remaining on orbits with  $q^*>30$ au and  $a^*>60$ au after 4 Gyr of evolution.  The fraction of objects with $a^*>150$ au is small in the cases of $M_d =15M_E$ and $M_d =20M_E$. This is not consistent with observations. But for $M_d =40M_E$, a third of the surviving objects have $a^*>150$ au.  There are Sedna-type objects with $q^* \ga 80$ au in this case.

\begin{figure}[h]
   \centering
   \resizebox{\hsize}{!}{\includegraphics{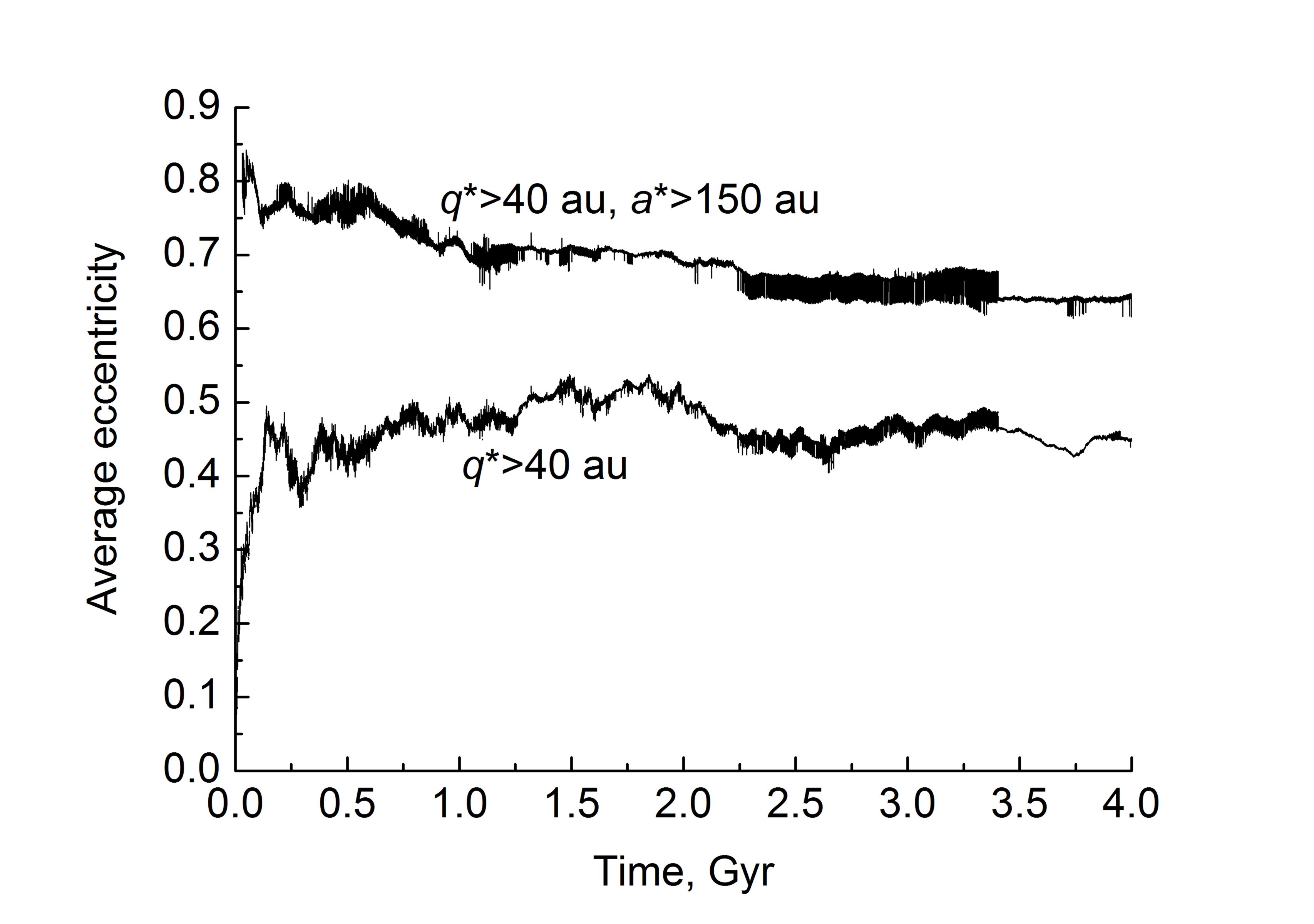}}
      \caption{ Evolution of average eccentricities for two datasets in the case with $M_d =40M_E$. The data are given with an interval of 200 thousand years. At the initial stage of evolution planetesimals are scattered by Neptune to high-eccentricity orbits. The first object appears in the region $q^*>40$ au, $a^*>150$ au at $t=26.8$ Myr.}
         \label{fig3}
   \end{figure}

Figure~\ref{fig3} shows the evolution of the average  eccentricities $\overline{e}$ in the regions $q^*>40$ au (no restriction on $a$) and $q^*>40$ au, $a^*>150$ au in the case with $M_d =40M_E$ . At the initial stage of evolution planetesimals are scattered by Neptune to high-eccentricity orbits. The first object appears in the region $q^*>40$ au, $a^*>150$ au at $t= 26.8$ Myr. We discuss the initial stage of evolution in more detail in Appendix C.   Then, there is a systematic decrease of $\overline{e}$ in the region  $q^*>40$ au, $a^*>150$ au due to the disk's self-gravity, and such an effect does not exist for smaller values of semimajor axis. This decrease is enough to increase perihelion distances of distant objects substantially and even to create Sedna-type objects. As shown in Appendix D, the significant changes in perihelion distances are associated with secular resonances between planetesimals.

Figure~\ref{fig2} shows a lack of objects with $a^*>500$ au in our model. This can be connected with the limit of 1000 au for semimajor axes in the basic integrations. We have checked this by additional integrations with the upper limit $a=5000$ au.
 
\subsection{Integrations with a limit value of 5000 au for semimajor axes}

We recomputed the variants discussed above, adopting the same initial conditions, but integrations were stopped when $a>5000$ au instead of 1000 au. We are aware that in this case galactic and stellar perturbations can be significant (we ignore them in this study), and we regard these computations only as a first attempt to estimate the influence of massive planetesimals in the inner Oort cloud.

\begin{figure}[h]
   \centering
   \resizebox{\hsize}{!}{\includegraphics{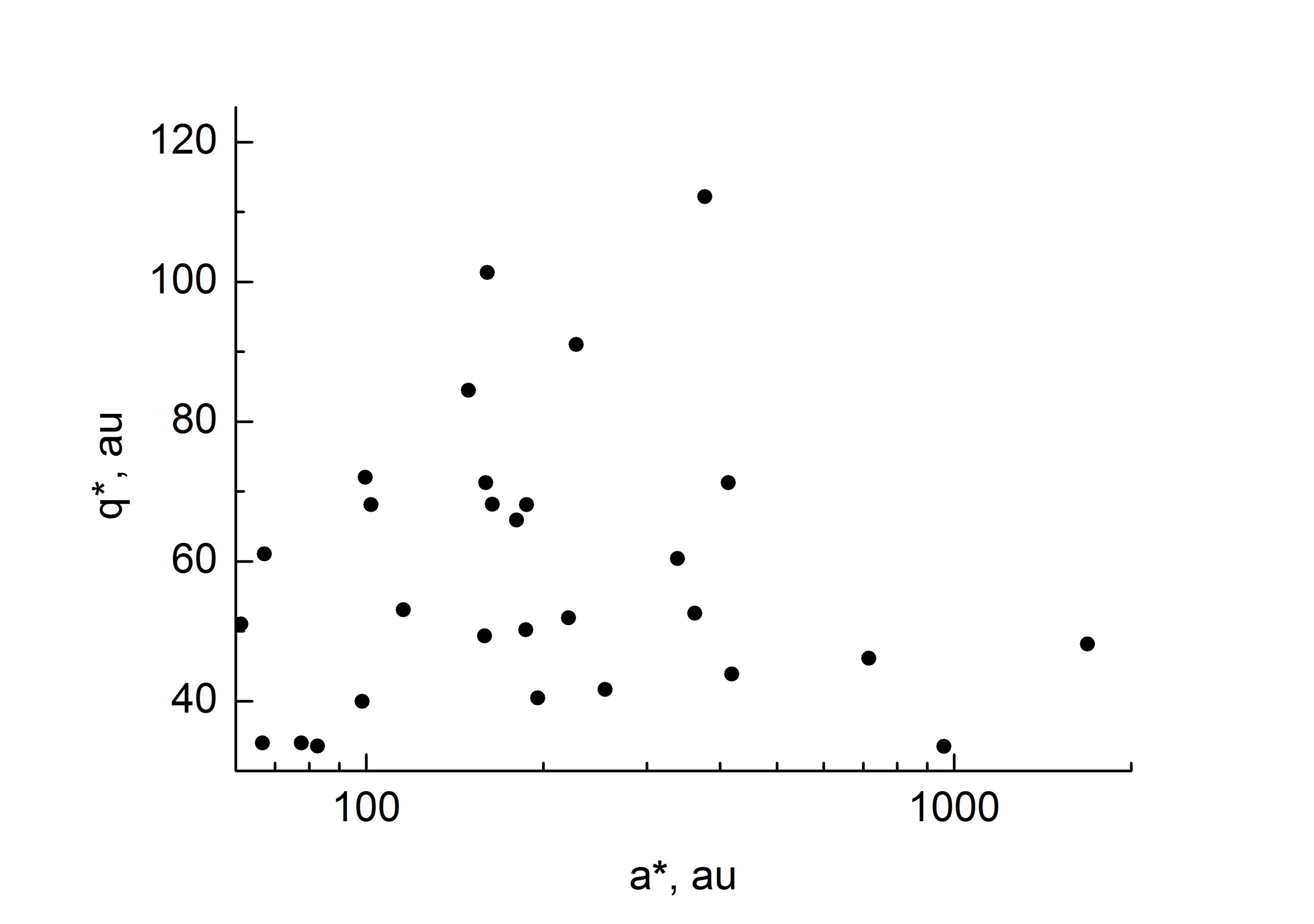}}
      \caption{Distribution of $a^*$ and $q^*$ after 4 Gyr for $M_d =40M_E$. }
         \label{fig4}
   \end{figure}

For the final distribution of orbital elements, we have not found major changes in comparison with data shown in Fig.~\ref{fig2}. Figure~\ref{fig4} shows the distribution of $a^*$ and $q^*$ for objects  remaining after 4 Gyr in the variant with $M_d =40M_E$. Again the majority of objects that survive for the age of the Solar System have   $q^*>40$ au. With the exception of two objects, all the others have $a^*<500$ au. The emergence of the object with $q^*=48.2$ au and  $a^*=1684$ au in this simulation indicates the importance of future consideration of the role of galactic and stellar perturbations in the formation of a population of distant TNOs.

\subsection{Distribution of angular orbital elements}

The observed clustering in arguments of perihelion, longitudes of ascending node and longitudes of perihelion 
$\pi$ for distant TNOs is the most controversial. \citet{TS14} first suggested that there is a concentration of arguments of perihelion near the value $\omega = 0^\circ$.  Later, \citet{BB16} noted that, to a greater extent, there is a grouping of longitudes of ascending nodes and  longitudes of perihelion.  On the one hand, many authors insist that this is due to observational biases \citep[e.g.,][]{Laetal17, Shetal17,Beetal20,CK20,Tr20,Naetal21}. On the other hand, \citet{Br17}, \citet{BB19}, and \citet{BrB21} have made the argument that after taking account of observational biases, clustering remains significant at the 99.6 percent confidence level.
 
\begin{figure}[h]
   \centering
   \resizebox{\hsize}{!}{\includegraphics{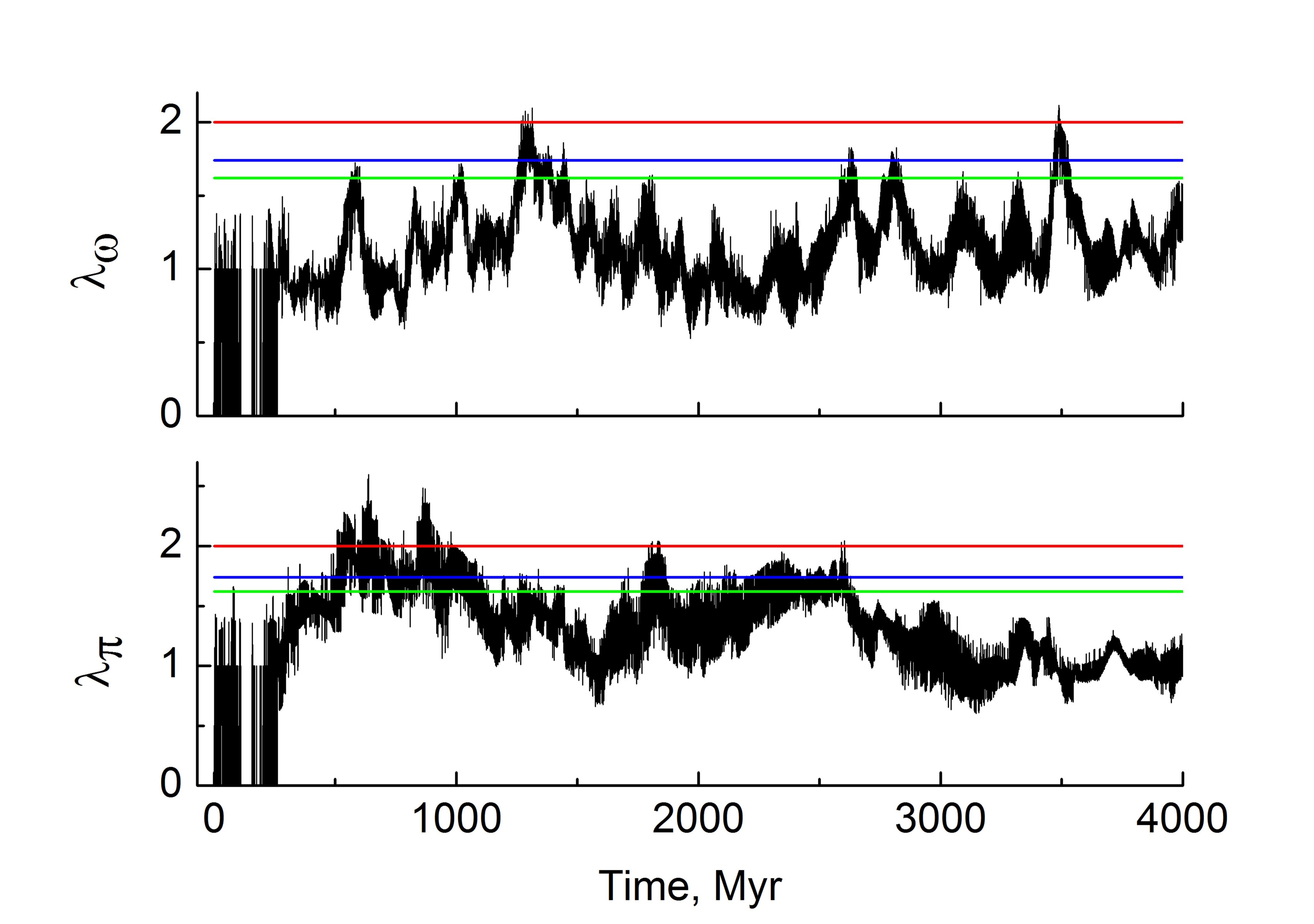}}
      \caption{Evolution of $\lambda_\pi$ and $\lambda_\omega$ for objects with $q^*>40$ au, $a^*>150$ au in the case with $M_d=40M_E$.  The data are given with an interval of 200 thousand years. The horizontal colored lines show the values  of 
$\lambda$  corresponding to the probabilities  $p=0.99$ (red line), $p=0.95$ (blue line),  $p=0.9$ (green line) that the hypothesis of a uniform distribution is rejected. }
         \label{fig5}
   \end{figure}

To characterize the degree of deviation of the orbital distribution in our simulations from a uniform distribution, we used the analogue of the  Kuiper statistic $\lambda=(D^+ - D^-)\sqrt{n}$, where $D^+$ and $D^-$ are the largest and smallest differences between the cumulative functions for the simulated distribution and the uniform distribution across all possible values, $n$ is the number of objects. Here, we give examples of the evolution of this quantity for longitudes of perihelion ($\lambda_\pi$) and arguments of perihelion  ($\lambda_\omega$) of objects with $q^*>40$ au, $a^*>150$ au in the case with $M_d=40M_E$ discussed in the previous subsection. Figure~\ref{fig5} shows $\lambda_\pi$  and $\lambda_\omega$ as a function of time.

The behavior of $\lambda_\pi$  and $\lambda_\omega$ is complicated. There are times with both large and small values of     
$\lambda_\pi$  and $\lambda_\omega$ .  We note that  according to the  Kuiper statistical test, the hypothesis of a uniform distribution is rejected with asymptotic probabilities of $p=0.99, 0.95$ and $0.9$ at $\lambda=2.00, 1.74$, and $1.62$, respectively. These values of $\lambda$ are shown in Fig.~\ref{fig5} as horizontal colored lines. For comparison, the observed set of 18 multiple-opposition TNOs with $q>40$ au, $150<a<1000$ au has $\lambda_\pi=1.83$ and  $\lambda_\omega=1.76$ (ignoring observational biases). Examples of the distribution of $\pi$ at different values of  $\lambda_\pi$ in our simulations are shown in Appendix E.

Figure~\ref{fig6} shows the distribution of $a^*$ and inclinations $i$ for the simulated objects  with $40<q^*<80$ au, $150<a^*<1000$ au remaining after 4 Gyr in the same variant  with $M_d =40M_E$. The average value of $i$ for the simulated objects is $19^\circ$, and it equals $18^\circ$ for the observed multiple-opposition TNOs with $40<q<80$ au, $150<a<1000$ au (the observable region of distant TNOs).
 
\begin{figure}[h]
   \centering
   \resizebox{\hsize}{!}{\includegraphics{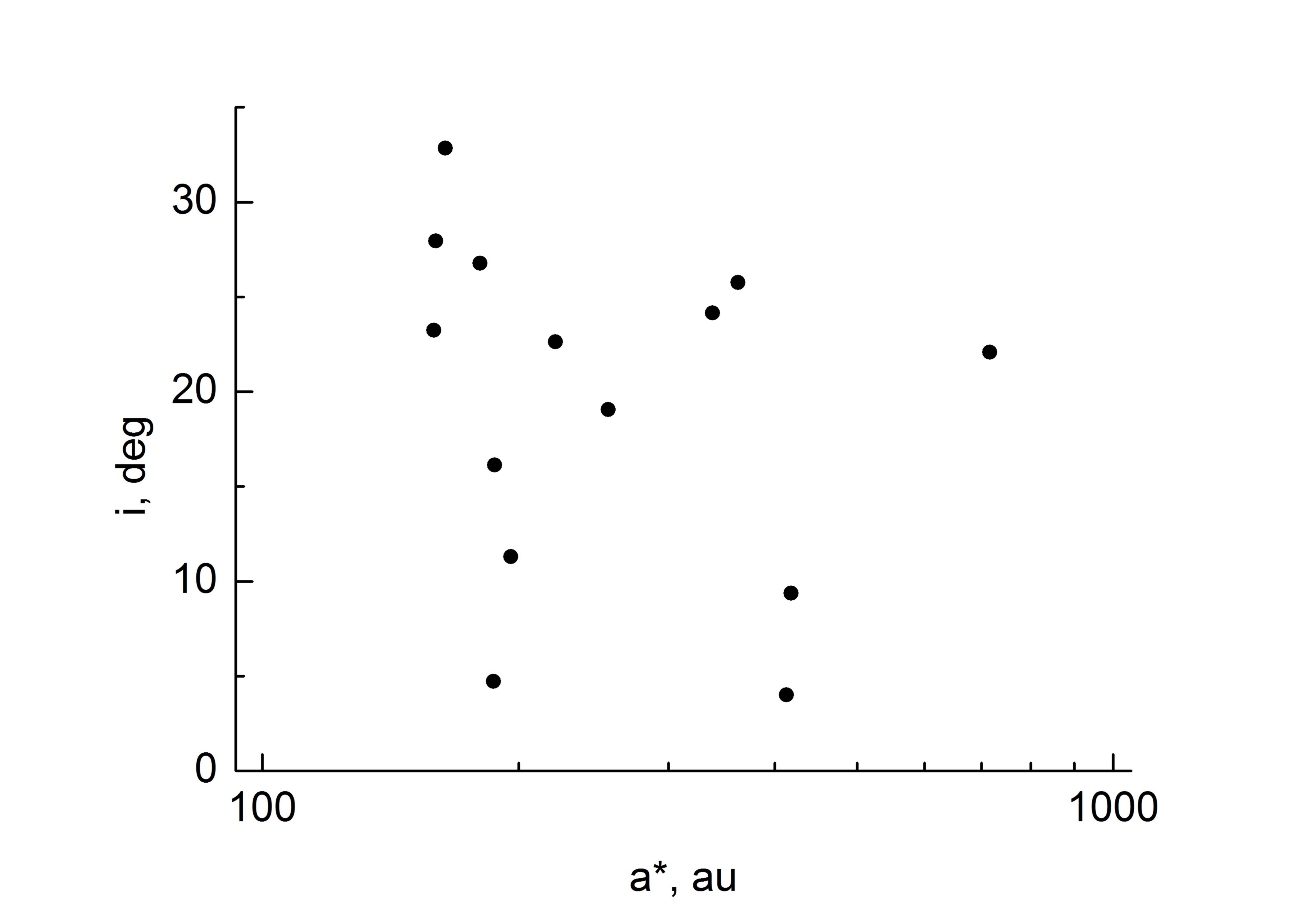}}
      \caption{Distribution of $a^*$ and $i$ for objects  with $40<q^*<80$ au, $150<a^*<1000$ au after 4 Gyr for  $M_d=40M_E$.}
         \label{fig6}
   \end{figure} 
 
\section{Discussion}

Our simulations show that the combined action of perturbations from the giant planets and self-gravity of the planetesimal disk produces distant TNOs in a wide range of the initial disk mass $M_d$. The majority of TNOs survive in the region $q>40$ au.
In this region, there is a tendency to slow decrease in eccentricities and increase in perihelion distances for objects with large semimajor axes ($a>150$ au). Thus, our model predicts the presence of a large population of Sedna-type objects. The fraction of TNOs with $q>40$ au, $a>150$ au surviving for the age of the Solar System depends on $M_d$. Our results favor $M_d$ greater than the value of $20M_E$ estimated in \citet{NM12} and \citet{NV16}. However, this investigation is only a preliminary effort aimed at a self-consistent study of the evolution of TNOs for the age of the Solar System and it has limitations.

First, our investigation uses initial conditions that are similar to those adopted in \citep{KS16,Neetal16}. In fact, these works are associated with studying a last non-catastrophic stage in the Nice model. In this model, various schemes were discussed, including studies of both the planetary instability stage and the pre-instability stage 
\citep[e.g.,][]{Tsetal05,Moetal07,BB10,Leetal11,NM12,Reetal15,QK19,DeSetal20,Cletal21,Moetal21}.  \citet{FB17} studied one case from \citep{Baetal11} that included a stage of the planetary instability and they compared simulations with a self-gravitating planetesimal disk and with a non-self-gravitating disk. The authors concluded that the orbital clustering observed in the distant trans-Neptunian region is unlikely to have a self-gravitational origin. However, the results were presented in this paper only for 20--35 Myr, and the objects with $q>40$ au, $a>150$ au were not analyzed separately. In our simulations, a noticeable secular change in orbits with $q>40$ au, $a>150$ au that is due to self-gravity appears after several hundred million years. Although we do not expect significant changes in our qualitative conclusions, we plan to consider their details for various schemes in future works.

The second caveat involves a relatively small number of massive objects. In our simulations, each massive planetesimal has a mass ranging from $0.03 M_E$ to $0.24 M_E$. These objects are much larger  than Pluto-mass bodies considered in \citep{Leetal11,NV16}. On the other hand, \citet{FI96} noted that objects of masses 1--5$\, M_E$ are always produced along with Uranus and Neptune in their simulations. Recent numerical experiments of \citet{Moetal21} indicated that it is plausible for $\sim 1\, M_E$ objects to emerge within the primordial trans-Neptunian region  on a $\sim10$ Myr timescale.  \citet{Zdetal20} tested the dependence of the "inclination instability'' effect on the number $N$ of massive objects in simulations. They found that the strength and duration of this effect increases with increasing $N$. In any case, further studies of the mass spectrum in the trans-Neptunian region are well motivated, especially with the discovery of new distant objects in the Solar System.

One more notable aspect concerns the possible action of galactic and stellar perturbations that are not included in our simulations. The synergy between these perturbations and self-gravity in the inner Oort cloud region is also an important topic for future investigations \citep[cf.,][]{Shetal19, BB21}.

\section{Conclusions} 

Distant TNOs are a natural consequence of models that include the migrating giant planets and a self-gravitating planetesimal disk. The majority of objects remaining in the trans-Neptunian region after 4 Gyr have $q>40$ au. Secular resonances between distant planetesimals play a major role in increasing their perihelion distances. This explains the origin of Sedna-type objects. In our integrations for the age of the Solar System, the degree of clustering of $\pi$ and $\omega$  for distant TNOs changes in a complicated way. The simulated distribution of inclinations after 4 Gyr of evolution and the observed distribution of inclinations have similar average values around $20^\circ$.

In the present paper, we do not attempt to fine tune our simulations to the observed distribution of TNOs. The real structure of the distant trans-Neptunian region is still very uncertain (only 18 multiple-opposition objects with $q>40$ au, $a>150$ au were known on  January 9, 2022). With an upcoming increase in detections of distant TNOs, further simulations, particularly those with more numerous massive bodies, and the inclusion of galactic and stellar perturbations will offer more rigorous constraints on the dynamical process discussed above. In this case, computations on time intervals on the order of the Solar System age will be a challenging problem, but such considerations provide intriguing possibilities for gaining a deeper understanding of the origins of the outer Solar System.

\begin{acknowledgements}
The author acknowledges the support of Ministry of Science and Higher Education of the Russian Federation under the grant 075-15-2020-780 (N13.1902.21.0039). The calculations were carried out using the MVS-10P supercomputer of the Joint Supercomputer Center of the Russian Academy of Sciences. The author thanks an anonymous referee for helpful comments.  
\end{acknowledgements}

%
%

\begin{appendix}
\section{Neptune's migration}
\begin{figure}[h]
   \centering
   \resizebox{\hsize}{!}{\includegraphics{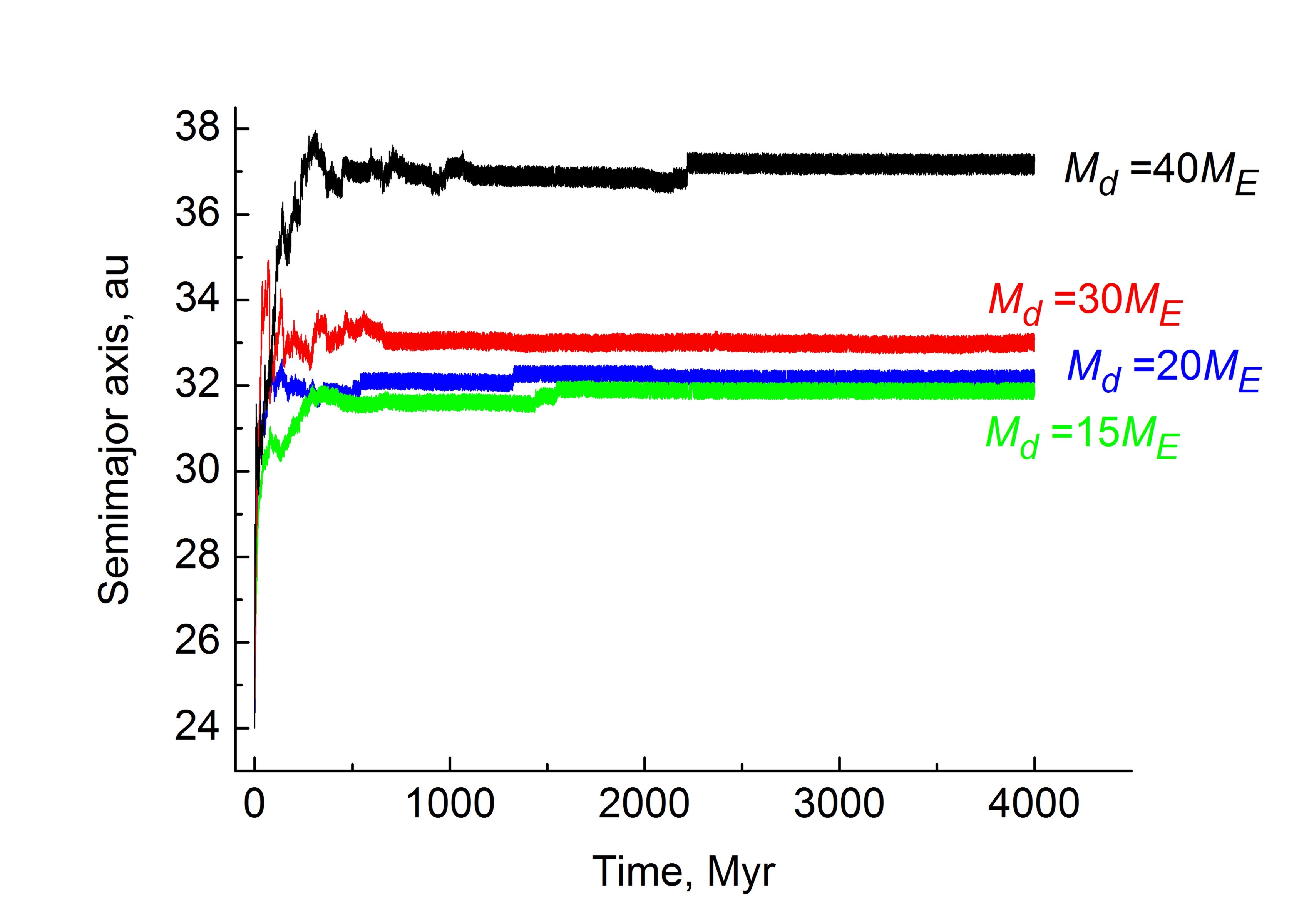}}
      \caption{Neptune's semi-major axis evolution for planetesimal disks with
$M_d =15M_E$ (green line), $M_d =20M_E$ (blue line), $M_d =30M_E$ (red line), and $M_d =40M_E$ (black line).}
         \label{figa}
   \end{figure}

Figure~\ref{figa} shows Neptune's semi-major axis evolution for planetesimal disks with
$M_d =15M_E$, $M_d =20M_E$, $M_d =30M_E$, and $M_d =40M_E$.
 
\section{Evolution of disk masses}
\begin{figure}[h]
   \centering
   \resizebox{\hsize}{!}{\includegraphics{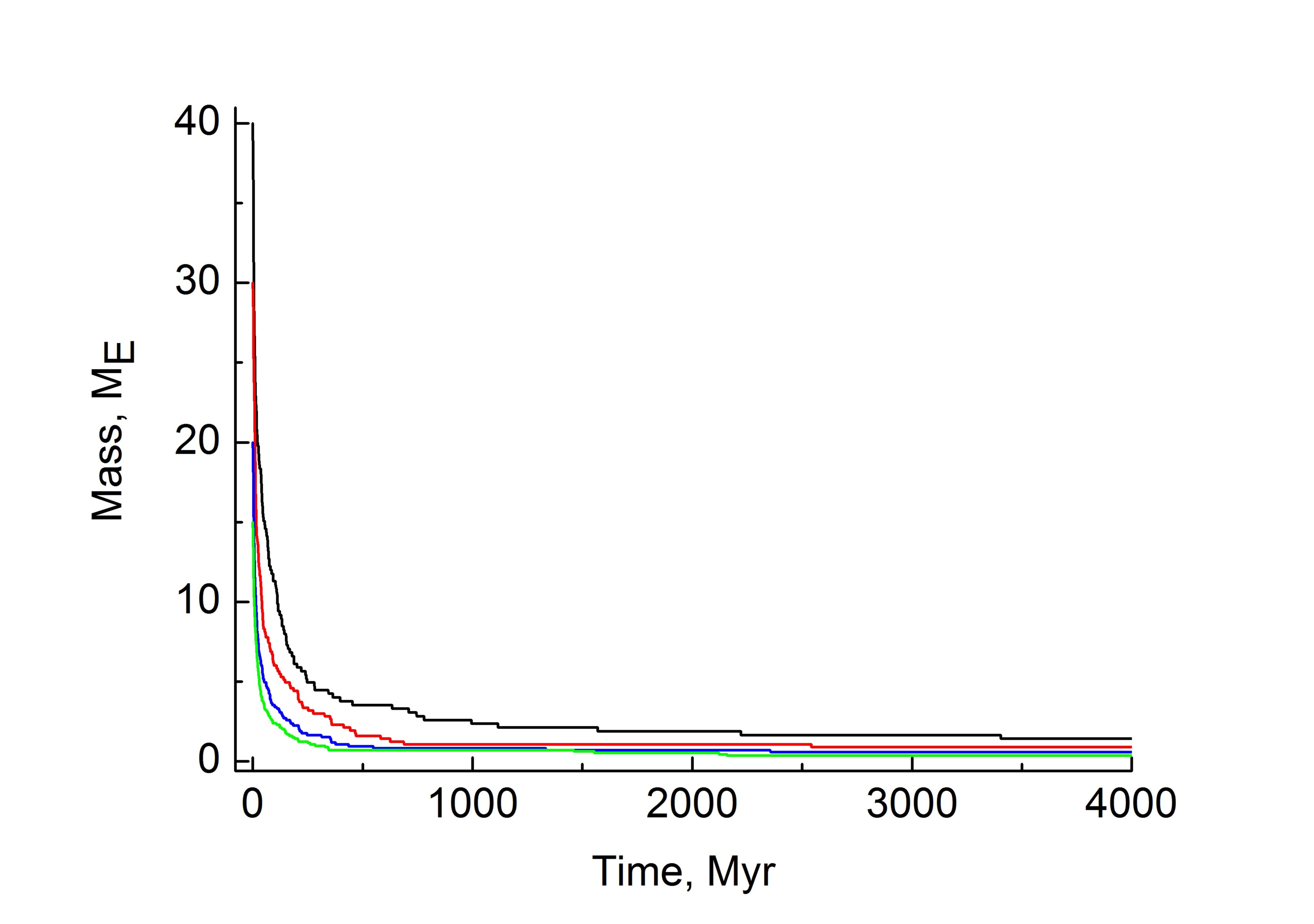}}
      \caption{Evolution of the disk mass for 
$M_d =15M_E$ (green line), $M_d =20M_E$ (blue line), $M_d =30M_E$ (red line), and $M_d =40M_E$ (black line).}
         \label{figb}
   \end{figure}

Figure~\ref{figb} shows the evolution of the disk mass for $M_d =15M_E$, $M_d =20M_E$, $M_d =30M_E$, and $M_d =40M_E$.

\section{Initial stage of evolution}

\begin{figure}[h]
   \centering
   \resizebox{\hsize}{!}{\includegraphics{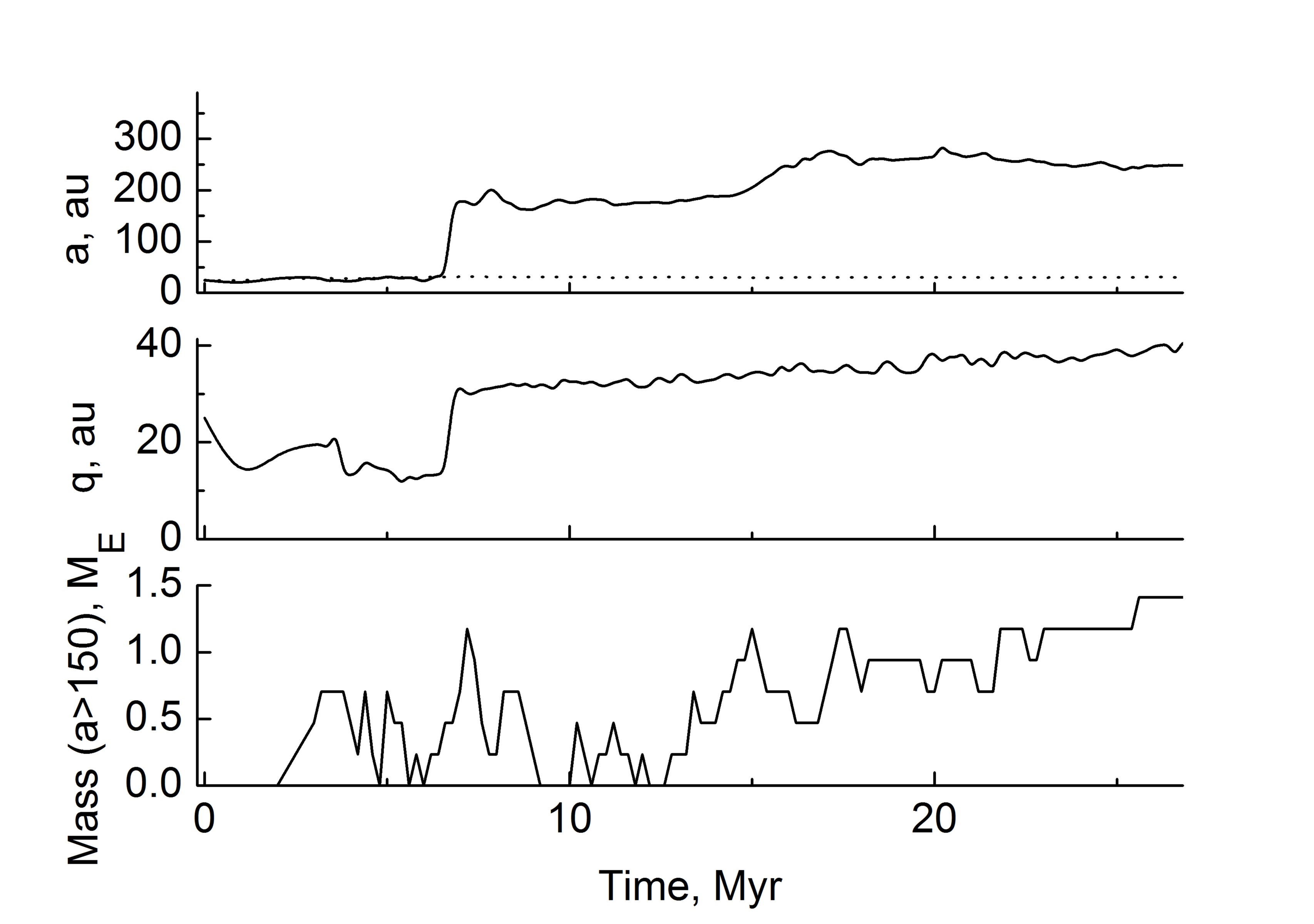}}
      \caption{Changes of the semimajor axis $a$ and the perihelion distance $q$ for the object reaching the region  $q^*>40$ au, $a^*>150$ au first (the top and middle panels). The dot line in the top panel corresponds to Neptune's semimajor axis. The bottom panel shows the total mass of planetesimals with $a>150$ au.}
         \label{figc}
   \end{figure}

Here, we discuss in more detail the initial stage of simulations shown in Fig.~\ref{fig3} for  $M_d =40M_E$. During this stage, planetesimals are scattered by Neptune to high-eccentricity orbits. Figure~\ref{figc} shows an example of the orbital evolution for the massless object reaching the region  $q^*>40$ au, $a^*>150$ au first. This object is ejected into orbit with $a=172$ au at $t=6.8$ Myr. At this time, a number of massive objects have already reached orbits with  $a>150$, as shown in the bottom panel of Fig.~\ref{figc}. Then the perihelion distance increases gradually, reaching 40 au at $t=26.8$ Myr. The semimajor axis of Neptune is equal to 30 au at this moment.

\section{Examples of the long-term orbital evolution}

\begin{figure}[h] 
   \centering
   \resizebox{\hsize}{!}{\includegraphics{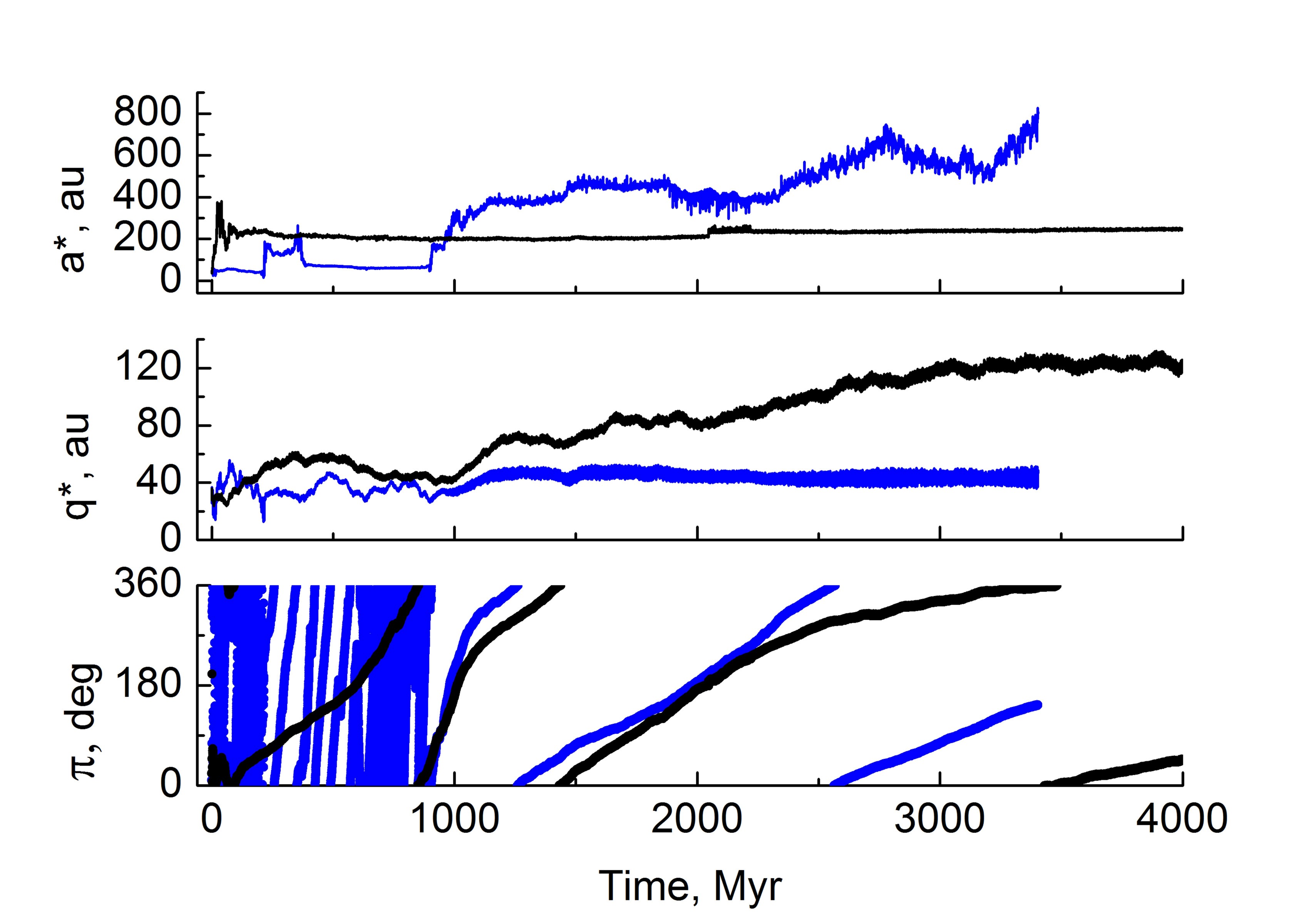}}
      \caption{Changes in $a^*$, $q^*$, and $\pi$ for a massive planetesimal (blue line) and a massless planetesimal (black line). The latter  survives  in the trans-Neptunian region for the age of the Solar System.}
         \label{figd1} 
   \end{figure}

Here, we describe typical features of the evolution of planetesimals into orbits with large perihelion distances.
Figure~\ref{figd1} shows the orbital evolution of  a massive planetesimal (blue line) and a massless planetesimal (black line). From 
$t= 0.9$ Gyr, both objects have $a^*>150$ au , and the rates of changes of $\pi$ for these objects are almost the same. Thus, these objects move in a secular resonance. This secular resonance drives the perihelion distance of the massless planetesimal up to  $q^*=132$ au. Such an orbit is located in the "inert zone" \citep{Saetal19,Sa20}, where changes in semimajor axes and perihelion distances over the Solar System age due to planetary perturbations are very small. The massive planetesimal is ejected by Neptune to   $a>1000$ au at $t= 3.4$ Gyr. But the massless planetesimal remains in the distant trans-Neptunian region until the end of simulations.

\begin{figure}[h]
   \centering
   \resizebox{\hsize}{!}{\includegraphics{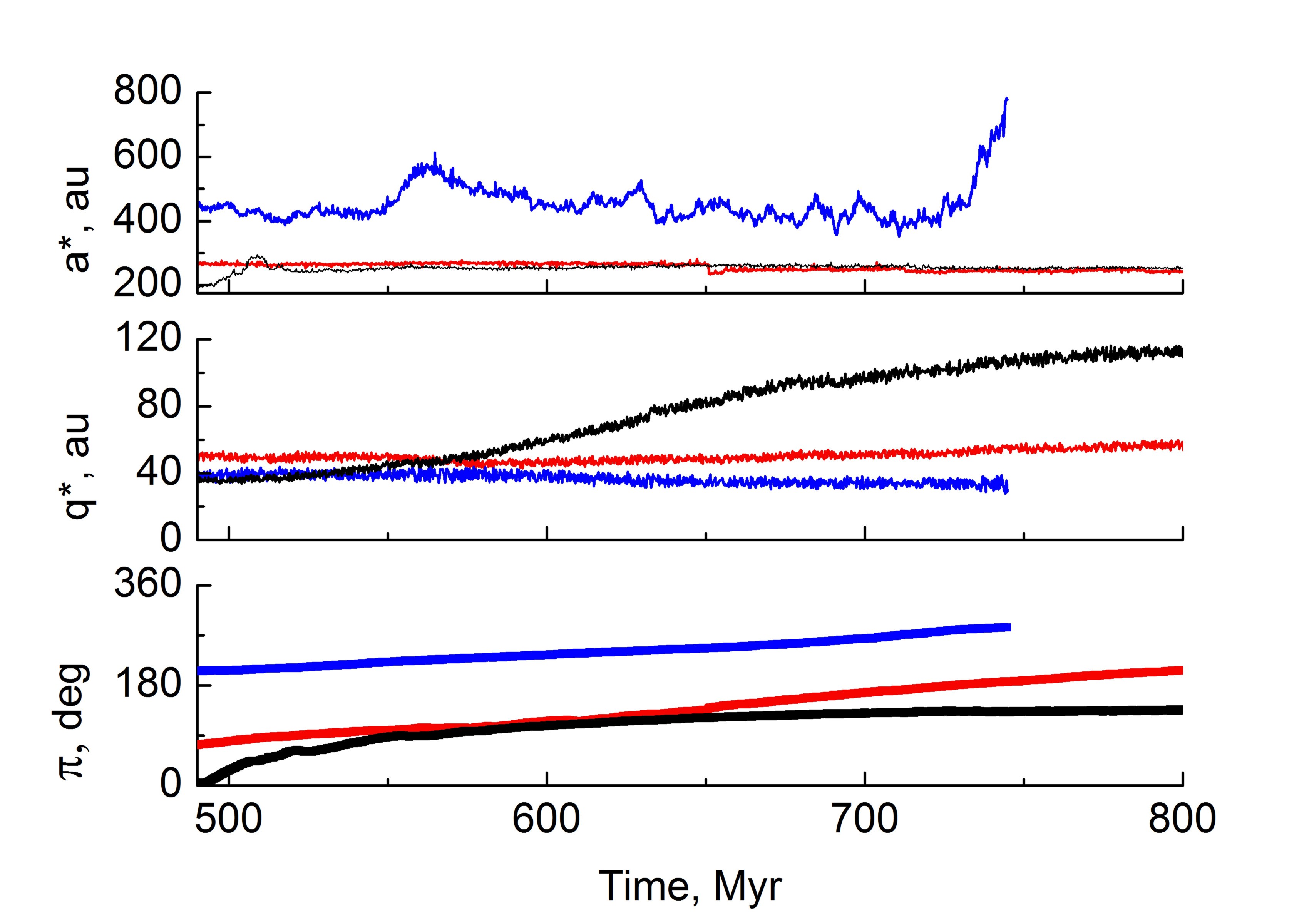}}
      \caption{Changes of $a^*$, $q^*$, and $\pi$ for two massive planetesimals with $a^*>150$ (blue and red lines) and a massless planetesimal (black line).}
         \label{figd2}
   \end{figure}

Figure~\ref{figd2}  shows an example of the orbital evolution where two massive planetesimals with $a^*>150$ au (blue and red lines) have similar precession rates of the line of apses. These objects move in a secular resonance until $t=0.75$ Gyr when one planetesimal is ejected to $a>1000$ au by Neptune. Near $t=0.52$ Gyr the massless planetesimal is also drawn in this resonance. 
As a result, the value of $q^*$ for the massless planetesimal changes from 35 au to 116 au. 

\section{Clustering of longitudes of perihelion}
\begin{figure}[h]
   \centering
   \resizebox{\hsize}{!}{\includegraphics{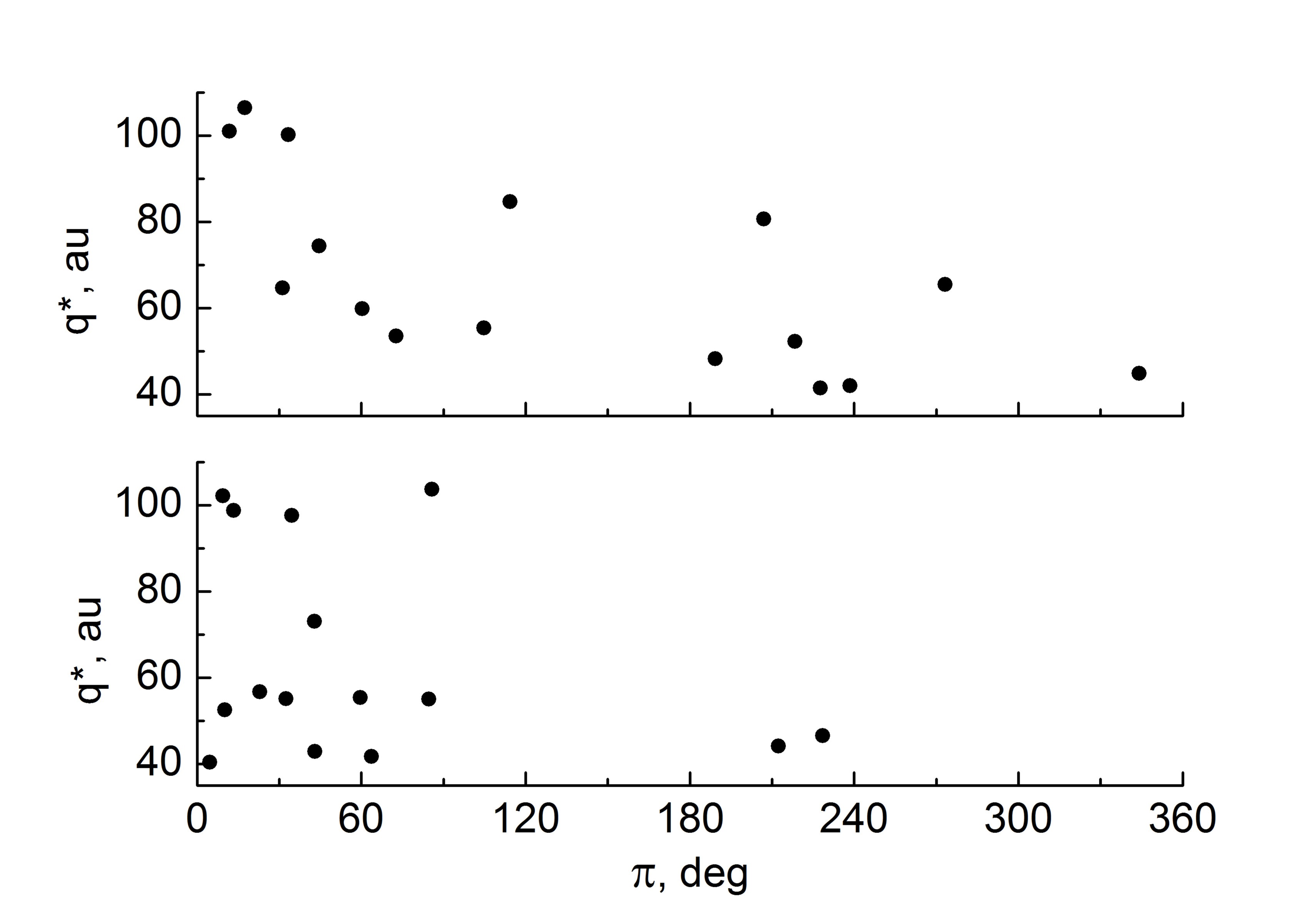}}
      \caption{ Distributions of $\pi$ and $q^*$ corresponding to  $\lambda_\pi=2.48$ at $t=862.4$ Myr (the bottom panel) and $\lambda_\pi=1.11$ at $t=953.6$ Myr (the top panel) in Fig.~\ref{fig5}.}
         \label{fige}
   \end{figure}

As shown in Fig.~\ref{fig5}, the degree of clustering of $\pi$ and $\omega$  for objects with $q^*>40$ au, $a^*>150$ au changes in an irregular way. We demonstrate in  Fig.~\ref{fige} the distributions of $\pi$ and $q^*$ corresponding to  $\lambda_\pi=2.48$ at $t=862.4$ Myr (strong clustering) and $\lambda_\pi=1.11$ at $t=953.6$ Myr (more uniform distribution) in Fig.~\ref{fig5}. 

\end{appendix}
\end{document}